# Weighted Indices for Evaluating the Quality of Research with Multiple Authorship

Ash Mohammad Abbas
Department of Computer Engineering
Aligarh Muslim University
Aligarh - 202002, India
Email: am.abbas.ce@amu.ac.in

*Abstract*—Devising an index to measure the quality of research is a challenging task. In this paper, we propose a set of indices to evaluate the quality of research produced by an author. Our indices utilize a policy that assigns the weights to multiple authors of a paper. We have considered two weight assignment policies: *positionally weighted* and *equally weighted*. We propose two classes of weighted indices: *weighted h-indices* and *weighted citation h-cuts*. Further, we compare our weighted h-indices with the original *h-index* for a selected set of authors. As opposed to h-index, our weighted h-indices take into account the weighted contributions of individual authors in multi-authored papers, and may serve as an improvement over h-index. The other class of weighted indices that we call *weighted citation h-cuts* take into account the number of citations that are in excess of those required to compute the index, and may serve as a supplement to h-index or its variants.

## I. INTRODUCTION

In many situations, it becomes necessary to assess the quality of research published by an author e.g. at the time of recruitment, and at the time of allocation of grants to carry out further research. To assess the quality of research produced by a researcher, one needs an index which should be able to quantify the quality of papers published by the author. However, the design of an index for an assessment of the quality of research produced by an author is a challenging task. The challenge comes from the fact that an index should be appropriate in the sense that it should be able to incorporate several other things beyond merely an authorship such as individual contributions in a collaborative work product.

To assess the quality of a paper and/or research produced by an author, a number of measures are proposed in the literature. For example, one such measure is, how many times the published paper is cited by other papers. This measure is called as the *citations count* for the particular paper. A researcher might have published a number of papers, therefore, the citation count for the researcher is the summation of the citation counts of his/her papers. The measure *citation count* is easy to compute. Specifically, one can compute citation count of an author using indexing databases such as Scopus, Web of Science, Google Scholar, etc. However, citation count may not be a good measure to assess the quality of research produced by a researcher as it does not take into account the number of authors and their individual contribution in the paper.

Another measure of the quality of research is $h$-index [1]. Axiomatic characterizations of $h$-index is carried out in [2], [3], [5]. The predictive power of $h$-index is discussed in [8], and its monotonicity is discussed in [4]. There are different views of peer researchers about the $h$-index— many of them consider $h$-index to be a good measure of the quality of research produced by an author [13], [14], [15], yet some authors are of the view that although $h$-index provides a ranking of authors, however, citations considered for determining the $h$-index need the context in which they are cited [6]. Some of the variations of $h$-index are also proposed in the literature such as rational or successive $h$-index [9], generalized $h$-index [10], age decaying $h$-index [11], and enhanced $h$-index [12]. An improvement over $h$-index called $g$-index is proposed in [16], [17].

The measure $h$-index is also based on the number of citation received by individual papers of an author, however, as opposed to the citation counts which sums out the number of citations of all $n$ cited papers of an author, $h$-index does not consider all cited papers of an author, it considers the papers with some minimum number of citations. Specifically, it considers top $h$ papers with at least $h$ citations for each individual paper, and the rest of the $n-h$ papers have less than or equal to $h$ citations. Like the citation count, the measure $h$-index is also easy to compute. However, similar to the citation count, it also does not take into account the number of authors of a paper and their individual contributions in the paper. It may happen that the top $h$-papers of a researcher that have at least $h$ citations may contain multiple authors in most of his/her papers. On the other hand, there can be another researcher who is the sole author of $h$ papers each of which is having at least $h$ citations. Although, both the researchers have the same $h$-index, however, there is a significant difference between the individual contributions of the two authors, and the $h$-index falls short of reflecting this difference.

It has been discussed in [27] that there should be well defined credits for co-authorship in the field of *computer science*, and many suggestion are made pertaining to the future of publishing in computer science and measures to address ethical issues. Approximately, two and half decades before the proposal of $h$-index, it was suggested in [24] that co-authors of a paper can be allocated harmonic weights to determine their relative contributions. It has been discussed in [23], [25], and [26] that harmonic weights can enable one to equitably share the authorship in multi-authored papers. To incorporate the effect of multiple authors in $h$-index, several modifications are proposed in the literature including modified $h$-index ($h_m$-index) [19], adaptive pure $h$-index ($h_a$) [21], fractional $h$-index ($h_f$-index) [22], $\hbar$-index [28]. We shall discuss in the later part of this paper that there is a need of some other index to supplement the $h$-index including its variants to clearly distinguish among the qualities of publications produced by different authors.

In this paper, we wish to answer the following research question: What is the impact of the number of authors and the position of a given author in the paper on the quality of research produced by the author? Is $h$-index in its original form sufficient to provide the quality of research or should it be modified to reflect the contributions of a given author in multi-authored papers? In this paper, we propose an index for quantifying the quality of research produced by an author. Our index takes into account the number of authors of the paper and tries to incorporate their individual contributions to the paper using the order of its authors. We discuss schemes for assigning weights to authors and analyze the impact of author positions on the proposed index. We compare the proposed index with the existing $h$-index and discuss their relative merits and demerits.

The rest of the paper is organized as follows. In Section II, we present the proposed weighted indices. Section III contains an analysis of the positional weights used in the proposed weighted indices. In Section IV, we present results and discussion. Section V contains a comparison of our work with the related works. Finally, the last section is for conclusion and future directions.

## II. PROPOSED INDICES

In this section, we wish to propose indices to evaluate the quality of research produced by a given author. The papers can either be written by a single author or multiple authors. As far as, a paper with a single author is concerned, all citations fall in the account of the sole author of the paper. However, the paper that has multiple authors, the number of citations received do not belong to only one author, but should be shared by all authors of the paper. Specifically, citations related to a multi-authored paper should be divided among all authors of the paper, preferably, according to their contributions. However, there is no mechanism that can exactly tell the individual contributions of the authors in a multi-authored paper. In the absence of such an exact mechanism, we assume that the position of an author in the list of the authors of the paper gives an indication about the contribution of the author to the paper, unless stated explicitly[1]. The assumption seems to be realistic. Generally, the author whose name appears as the first author has the maximum contribution, and the contribution of the second author is less than that of the first author, and so on. Therefore, one can assume that the position of an author in the list of authors of the paper provides an indication about the contributions of the author to the paper.

---

[1]Generally, it is a hidden assumption that the order of authors is the order of their individual contribution in decreasing order, unless stated otherwise. Sometimes, authors provide a footnote that the names of authors in the list are in alphabetical order. In that case, the order of authors in the list does not imply the order of their contributions, and it becomes difficult to determine their order of contributions. However, in such an exceptional case, one may assign equal weights to all authors of the paper.

We now discuss the notion of *weighted citations*, which shall be used to define the weighted indices. Let $c_i$ be number of citations of $i$th paper of an author, the *weighted citations* of the $i$th paper of the author are as follows.

$$\psi = c_i w_i \qquad (1)$$

where $w_i$ is the weight assigned to the given author for his/her $i$th paper under a weight assignment scheme, say $\mathcal{A}$. In the next section, we describe two weight assignment schemes, namely, *equal weight assignment scheme* and *positional weight assignment scheme*. Using the notion of weighted citations, we define an index that we call *weighted citation aggregate*, as follows.

*Definition 1 (Weighted Citation Aggregate):* Let there be $n$ publications of a researcher (or an author), the *weighted citation aggregate*, $\psi$, of the given author is as follows.

$$\psi = \sum_{i=1}^{n} c_i w_i \qquad (2)$$

where, $c_i$ is the number of citations of the $i$th paper of the author, and $w_i$ is the weight of the author in his/her $i$th paper under a weight assignment scheme $\mathcal{A}$.

As mentioned above, there are two weight assignment schemes— positional, and equal. Accordingly, there are two weighted citation aggregates— *positionally weighted citation aggregate*, and *equally weighted citation aggregate*. Let $\psi_p$ be *positionally weighted citation aggregate* under a positional weight assignment scheme, say $\mathcal{P}$, and $\psi_e$ be *equally weighted citation aggregate* under *equal weight assignment scheme*, say $\mathcal{E}$. We have,

$$\psi_p = \sum_{i=1}^{n} c_i w_i, \forall i \quad w_i \in \mathcal{P} \qquad (3)$$

and,

$$\psi_e = \sum_{i=1}^{n} c_i w_i, \forall i \quad w_i \in \mathcal{E}. \qquad (4)$$

Note that our notion of the weighted citations can be used to modify the definition of $h$-index [1] so that it is able to incorporate the number of authors and/or position of the given author in the list of authors of the paper. Note that $h$-index is defined as the number such that $h$ papers out of $n$ cited papers of an author have received at least $h$ citations and $n-h$ papers have less than or equal to $h$ citations. In the following, we use the notion of weighted citations to propose an improvement over $h$-index.

*Definition 2 (Weighted h-index):* Let $c_i$ be the citations of $i$th paper of a given author, and let $w_i$ be the weight assigned to the author using a weight assignment policy, say $\mathcal{A}$. Let $n$ be the total number of papers published by the author and that are cited by other papers. The weighted $h$-index is the number such that weighted citations of the given author for his/her $h$ papers is at least $h$, and the remaining $n - h$ papers have at most $h$ citations. In other words,

$$h_w|_{w_i \in \mathcal{A}} = h_w, \text{if} \left( \min_{i=1}^{h_w}\{c_i w_i\} \geq h_w, \max_{j \neq i}\{c_j w_j\} \leq h_w \right). \quad (5)$$

Again, as mentioned earlier, there are two schemes for assigning the weights. As a result, there are two types of weighted $h$ indices: *positionally weighted $h$-index*, $h_p$, drawing their weights using policy, $\mathcal{P}$; and *equally weighted $h$-index*, $h_e$, drawing their weights using policy $\mathcal{E}$. In other words,

$$h_p|_{w_i \in \mathcal{P}} = h_p, \text{if} \left( \min_{i=1}^{h_p}\{c_i w_i\} \geq h_p, \max_{j \neq i}\{c_j w_j\} \leq h_p \right) \quad (6)$$

and,

$$h_p|_{w_i \in \mathcal{A}} = h_p, \text{if} \left( \min_{i=1}^{h_p}\{c_i w_i\} \geq h_p, \max_{j \neq i}\{c_j w_j\} \leq h_p \right). \quad (7)$$

Note that weighted citation aggregates defined above take into account *all* cited papers of an author. In what follows, we define an index that takes into account the weighted citations of papers that form the weighted $h$-core of the given author. We call this index *weighted citation $h$-cut*, whose definition is as follows.

*Definition 3 (Weighted Citation H-Cut):* Let $h_w$ be the weighted $h$-index of an author, and let $\mathbf{H_w}$ be the set of weighted citations of papers of a given author that contribute to the weighted $h$-index of the author. The *weighted citation $h$-cut* of the author, $\xi_w$, is as follows.

$$\xi_w = \sum_{i=1}^{|H_w|} c_i w_i \quad (8)$$

where, $|H_w|$ denotes the cardinality of the set $\mathbf{H_w}$.
We would like to mention again that there are two policies for assigning the weights, namely, positional ($\mathcal{P}$) and equal ($\mathcal{E}$). Accordingly, there are two *weighted citation $h$-cut*, which are as follows.

$$\xi_p = \sum_{i=1}^{|H_p|} c_i w_i|_{w_i \in \mathcal{P}} \quad (9)$$

$$\xi_e = \sum_{i=1}^{|H_e|} c_i w_i|_{w_i \in \mathcal{E}} \quad (10)$$

where, $|H_p|$ and $|H_e|$ are the cardinalities of the sets forming $h$-cores under policies $\mathcal{P}$ and $\mathcal{E}$, respectively.

In what follows, we describe how to assign the positional weights.

## III. ASSIGNMENT OF POSITIONAL WEIGHTS

The weight, $w_i$ of the author in the $i$th paper is related to his/her position in the list of authors of $i$th paper. We call it *positional weight* or the *contribution* of the author in his/her $i$th paper. The positional weights satisfy the following properties.

- For papers with multiple authors, the positional weights have to be designed in such a manner so that the first author is given more weight than the second author, the second author is given more weight than the third author, and so on. In other words, $w_i > w_j, \forall i < j$.

- The summation of these weights for all authors of the paper is equal to one. In other words, let there be $k$ authors a paper, and author $j$, $1 \leq j \leq k$, be assigned a weight $w_j$, then the following holds.

$$\sum_{j=1}^{k} w_j = 1. \quad (11)$$

There is a question: How can one assign these weights so that $w_i > w_j, \forall i < j$? In what follows, we present an intuitive scheme for assigning weights to authors of the same paper.

*Lemma 1:* Let there be $k$ authors of a paper, and $\mathbf{w} = \{w_j\}$ be the weights assigned to the $j$th author of the paper where $j$ varies from 1 to $k$. A possible scheme for assigning weights to $j$th author of the paper can be

$$w_j = \frac{2(k - j + 1)}{k(k + 1)} \quad (12)$$

where, $0 \leq w_j \leq 1$, and $\sum_{j=1}^{k} w_j = 1$.

*Proof:* Given that there are $k$ authors of the paper. Let $j$ be an integer, varying from 1 to $k$, that represents the position of an author in list of the authors of the paper. Then, summation of the positions of all authors of the paper is given by

$$S_p = \sum_{j=1}^{k} j$$
$$= \frac{k(k + 1)}{2}. \quad (13)$$

Let the contribution of an author whose name appears at the $j$th position in the list of authors of the paper be $\beta_j$ such that $0 \leq \beta_j \leq 1$, $\sum_{j=1}^{k} \beta_j = 1$, and $\beta_i > \beta_j$, $\forall i < j$. Intuitively, the position of authors and their contributions can be listed as follows.

| $j:$ | 1 | 2 | 3 | .. | $j$ | .. | $k$ |
|---|---|---|---|---|---|---|---|
| $\beta_j:$ | $\frac{k}{S_p}$ | $\frac{k-1}{S_p}$ | $\frac{k-2}{S_p}$ | .. | $\frac{k-j+1}{S_p}$ | .. | $\frac{1}{S_p}$. |

This is in contrast to the shares of persons whose ratios of shares are specified. The contrast lies in the fact that in case of shares the contribution of a person is more if his/her ratio is larger irrespective of his/her position in the ratios, however, the contribution of an author is assumed to be decreasing with the position of author in the list of authors of the paper.

The contribution of an author appearing at the $j$th position is given by

$$\beta_j = \frac{k - j + 1}{S_p}$$
$$= \frac{k - j + 1}{\frac{k(k+1)}{2}}$$
$$= \frac{2(k - j + 1)}{k(k + 1)}. \quad (14)$$

which is same as given by (12) except the name of the variable which is $w_j$ instead of $\beta_j$.

Formally, we prove (12) using the *principle of mathematical induction* as follows.

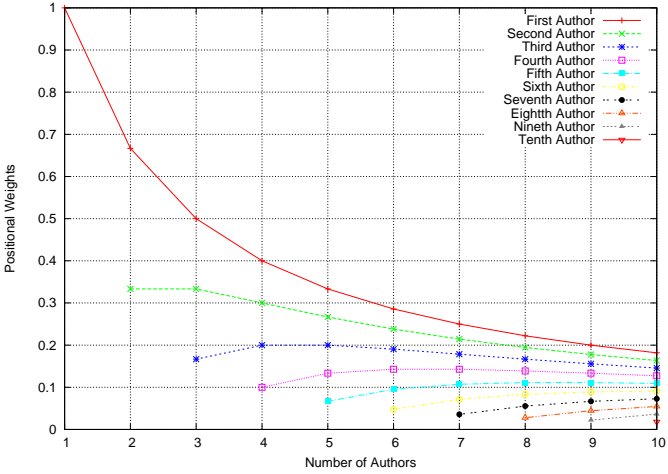

Fig. 1. Positional weights of authors as a function of the number of authors.

- For $k = 1$, $j$ can have only one value and that is equal to 1, therefore, $w_1 = 1$, which is true. As a result, (12) holds for $k = 1$.
- Assume that (12) is satisfied for an integer $k$, where $k$ is the number of authors of the paper. Adding one more author to the list of authors of the paper, the number of authors becomes $k+1$. For $k+1$ authors, we can rewrite $j$ and $\beta_j$ as follows.

  | $j$: | 1 | 2 | 3 | .. | $j$ | .. | $k+1$ |
  |---|---|---|---|---|---|---|---|
  | $\beta_j$: | $\frac{k+1}{S_p}$ | $\frac{k}{S_p}$ | $\frac{k-1}{S_p}$ | .. | $\frac{k+1-j+1}{S_p}$ | .. | $\frac{1}{S_p}$ |

  The value of $S_p$ is given by
  $$\begin{aligned} S_p &= S_p + (k+1) \\ &= \frac{k(k+1)}{2} + (k+1) \\ &= (k+1)\left(\frac{k}{2}+1\right) \\ &= \frac{(k+1)\{(k+1)+1\}}{2}. \end{aligned}$$

  The value of $\beta_j$ is given by
  $$\begin{aligned} \beta_j &= \frac{k+1-j+1}{\frac{(k+1)\{(k+1)+1\}}{2}} \\ &= \frac{2\{(k+1)-j+1\}}{(k+1)\{(k+1)+1\}}. \end{aligned}$$

  which is true (or satisfied for $k = k+1$). Therefore, by principle of mathematical induction (12) holds for all integers. ∎

Table I shows the number of authors and their weights according to their relative position in the list of authors of a paper. Note that the weights decrease with increasing the position number. Also, the weight of the same position decreases with the increase in the number of authors. By using unequal position based weights the weight of the author whose name figures out later in the list of authors of the paper is assigned less weight than the author whose name appears earlier.

Figure 1 shows positional weights of authors as a function of the number of authors of a paper. Note that the successive points lying from top to bottom along the vertical grid lines show the weights starting from author position *one* onwards. For example, if there are 7 authors, then the vertical grid line corresponding to the number of authors to be equal to 7 contains the weights of first, second, ..., seventh author from the top to bottom. The same is true for the remaining numbers of authors.

Based on the Lemma 1, we have the following corollary.

*Corollary 1:* Let there be $k$ authors of a paper. The difference in the weights assigned to the first author and the last author is a positive quantity for $k > 1$ and is given by

$$w_1^p - w_k^p = \frac{2}{k}\left(\frac{k-1}{k+1}\right). \quad (15)$$

*Proof:* Using (12), the weight of the first author is

$$w_1^p = \frac{2}{k+1}. \quad (16)$$

The weight of the $k$th author is

$$w_k^p = \frac{2}{k(k+1)}. \quad (17)$$

Therefore, the difference of the weights of the first and the last author is

$$\begin{aligned} w_1^p - w_k^p &= \frac{2}{k+1} - \frac{2}{k(k+1)} \\ &= \frac{2}{k}\left(\frac{k}{k+1} - \frac{1}{k+1}\right) \\ &= \frac{2}{k}\left(\frac{k-1}{k+1}\right). \end{aligned}$$

which is a positive quantity for $k > 1$, and 0 for $k = 1$[2]. ∎

Note that if all authors were assigned an equal weight, then the weight of each author would have been $\frac{1}{k}$. We now have the following lemma that gives the difference between the weights assigned under equal weights scheme and the weights assigned under position based scheme.

*Lemma 2:* Let $k$ be the number of authors of a paper. Let $w_j^e$ be the weight assigned to the $j$th author of the paper in equal weight scheme, and $w_j^p$ be the weight assigned to $j$th author in positional weight scheme. The amount of increase/decrease in the weight of the $j$th author in positional weight scheme as compared to equal weight scheme is given by the following expression.

$$w_j^p - w_j^e = \frac{1}{k} - \frac{2j}{k(k+1)}. \quad (18)$$

*Proof:* Using (12) weight assigned to $j$th author in the positional weight scheme is

$$w_j^p = \frac{2(k-j+1)}{k(k+1)}. \quad (19)$$

---

[2]Note that $k = 1$ means that there is only one author, who is the first author as well as the last author. As a result, there is no difference between the weights of the first author and the last author.

TABLE I
NUMBER OF AUTHORS AND THEIR POSITIONAL WEIGHTS.

| Number of Authors | $w_1$ | $w_2$ | $w_3$ | $w_4$ | $w_5$ | $w_6$ | $w_7$ | $w_8$ | $w_9$ | $w_{10}$ |
|---|---|---|---|---|---|---|---|---|---|---|
| 1 | 1 | | | | | | | | | |
| 2 | $\frac{2}{3}$ | $\frac{1}{3}$ | | | | | | | | |
| 3 | $\frac{3}{6}$ | $\frac{2}{6}$ | $\frac{1}{6}$ | | | | | | | |
| 4 | $\frac{4}{10}$ | $\frac{3}{10}$ | $\frac{2}{10}$ | $\frac{1}{10}$ | | | | | | |
| 5 | $\frac{5}{15}$ | $\frac{4}{15}$ | $\frac{3}{15}$ | $\frac{2}{15}$ | $\frac{1}{15}$ | | | | | |
| 6 | $\frac{6}{21}$ | $\frac{5}{21}$ | $\frac{4}{21}$ | $\frac{3}{21}$ | $\frac{2}{21}$ | $\frac{1}{21}$ | | | | |
| 7 | $\frac{7}{28}$ | $\frac{6}{28}$ | $\frac{5}{28}$ | $\frac{4}{28}$ | $\frac{3}{28}$ | $\frac{2}{28}$ | $\frac{1}{28}$ | | | |
| 8 | $\frac{8}{36}$ | $\frac{7}{36}$ | $\frac{6}{36}$ | $\frac{5}{36}$ | $\frac{4}{36}$ | $\frac{3}{36}$ | $\frac{2}{36}$ | $\frac{1}{36}$ | | |
| 9 | $\frac{9}{45}$ | $\frac{8}{45}$ | $\frac{7}{45}$ | $\frac{6}{45}$ | $\frac{5}{45}$ | $\frac{4}{45}$ | $\frac{3}{45}$ | $\frac{2}{45}$ | $\frac{1}{45}$ | |
| 10 | $\frac{10}{55}$ | $\frac{9}{55}$ | $\frac{8}{55}$ | $\frac{7}{55}$ | $\frac{6}{55}$ | $\frac{5}{55}$ | $\frac{4}{55}$ | $\frac{3}{55}$ | $\frac{2}{55}$ | $\frac{1}{55}$ |

The weight assigned to $j$th author under equal weight scheme does not depend upon the value of $j$ and is same for all authors of the paper. More precisely, its value is $\frac{1}{k}$. In other words,

$$w_j^e = \frac{1}{k}. \quad (20)$$

The difference of weights is given by

$$\begin{aligned} w_j^p - w_j^e &= \frac{2(k-j+1)}{k(k+1)} - \frac{1}{k} \\ &= \frac{2(k-j+1) - (k+1)}{k(k+1)} \\ &= \frac{k+1-2j}{k(k+1)} \\ &= \frac{1}{k} - \frac{2j}{k(k+1)}. \end{aligned}$$

∎

In the following, we provide another lemma that gives the difference between the weights of authors in the equal weight scheme and the positional weight scheme.

*Lemma 3:* Let $w_1^p$ and $w_k^p$ be the weights of the first and the last authors, respectively, in the position based weight assignment scheme, and let $w_e$ be the weight assigned to each author in equal weight assignment scheme. The following expression holds.

$$w_1^p - w_e = -(w_e - w_k^p) = \frac{1}{k}\left(\frac{k-1}{k+1}\right), \quad \forall k > 1. \quad (21)$$

*Proof:* Note that the weight assigned to each author in equal weight assignment scheme is given by

$$w_e = \frac{1}{k}. \quad (22)$$

Using (16) and (22), we have

$$\begin{aligned} w_1^p - w_e &= \frac{2}{k+1} - \frac{1}{k} \\ &= \frac{1}{k}\left(\frac{2k}{k+1} - 1\right) \\ &= \frac{1}{k}\left(\frac{2k-k-1}{k+1}\right) \\ &= \frac{1}{k}\left(\frac{k-1}{k+1}\right). \end{aligned} \quad (23)$$

Using (22) and (17), we get

$$\begin{aligned} w_e - w_k^p &= \frac{1}{k} - \frac{2}{k(k+1)} \\ &= \frac{1}{k}\left(1 - \frac{2}{k+1}\right) \\ &= \frac{1}{k}\left(\frac{k+1-2}{k+1}\right) \\ &= \frac{1}{k}\left(\frac{k-1}{k+1}\right). \end{aligned} \quad (24)$$

Reversing the sign of the above expression, we have

$$w_k^p - w_e = -\frac{1}{k}\left(\frac{k-1}{k+1}\right).$$

∎

In other words, weights assigned to each author in equal weight assignment scheme are lying on half way of the weights assigned to the first and the last authors in position based weight assignment scheme. This also indicates that the amount of increase in the positional weights of some of the authors is equal to the amount of decrease in the positional weights of the remaining authors.

Figure 2 shows the weight assigned to the first author and the last author of a paper as a function of the number of authors of a paper under positional weight scheme and under equal weight scheme. In case of positional weight scheme, we observe that when a person is the sole author of the paper, he/she is assigned a weight of 1. When the number of authors is two, the first author is assigned a weight of $\frac{2}{3} \approx 0.666$, and the second author is assigned a weight of $\frac{1}{3} \approx 0.333$. Note that if the two authors would have been assigned an equal weight, the weight of each of the author would have been $\frac{1}{2} = 0.5$. As compared to equal weight scheme, the increase in the weight of the first author under positional weight scheme is $16.6\%$; the same is the decrease in the weight of the second author. As mentioned earlier, if there are $k$ authors in a paper, the weight assigned to each authors is $\frac{1}{k}$ under equal weight scheme and weights assigned according to the position of authors is given by (12). The weight assigned under equal weight scheme fall in between the weights assigned to the first author and the last author of the paper.

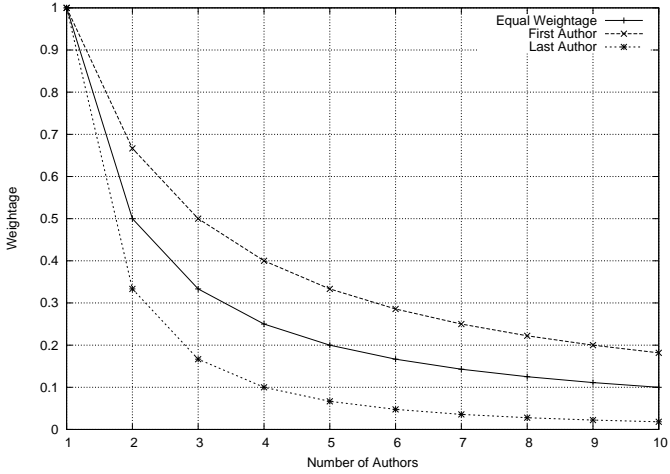

Fig. 2. Positional weights of the first and the last authors as a function of the number of authors.

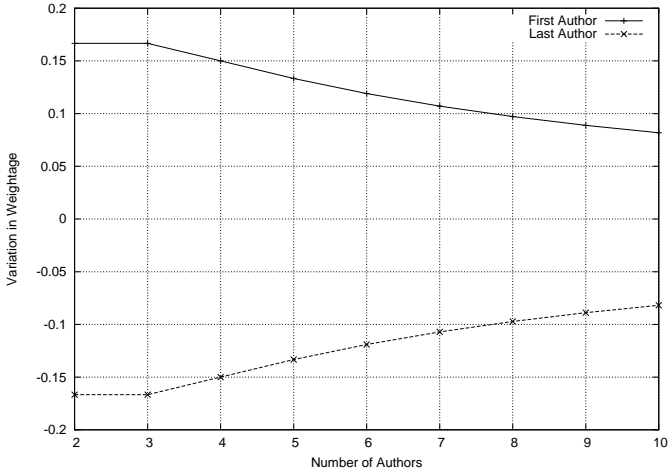

Fig. 3. Variations in the positional weights of the first and the last authors as a function of the number of authors.

Figure 3 shows the relative increase or decrease in the weights of the first author and the last author of a paper using positional weight scheme as compared to the weights assigned using equal weight scheme. The minus sign shows that it is a decrease in the weight under positional weight scheme as compared to weights under equal weight scheme. We observe that the weight of the first author of a paper is always increased under positional weight scheme with respect to equal weight scheme, and the weights of the second or later authors are always decreased under positional weight scheme as compared to equal weight scheme. Also, note that under positional weight scheme the increase in the weight of the first author is exactly the same as the decrease in the weight of the last author.

Note that under position based weight assignment scheme, the difference between the weights of the first and the last authors decreases with the number of authors (see Figure 2). Similarly, in Figure 3, the increase in the weight of the first author with respect to weight assigned to each author in equal weight assignment scheme reduces as a function of the number of authors, and the decrease in the weight of the last author under positional weight assignment scheme also reduces as compared to the weight assigned to each author using equal weight assignment scheme. This is in accordance with the limits given by the following lemma.

*Lemma 4:* Let $w_1^p$ and $w_k^p$ be the weights assigned to the first and the last authors, respectively, under position based weight assignment scheme, and let $w_e$ be the weight assigned to each author using equal weight assignment scheme. The following limits hold.

$$\lim_{k \to \infty} (w_1^p - w_k^p) = 0. \quad (25)$$

$$\lim_{k \to \infty} (w_1^p - w_e) = \lim_{k \to \infty} (w_e - w_k) = 0. \quad (26)$$

*Proof:* Using (16) and (17), we have

$$w_1^p - w_k^p = \frac{2}{k+1} - \frac{2}{k(k+1)}.$$

Taking the limits, we have

$$\begin{aligned}
\lim_{k \to \infty} (w_1^p - w_k^p) &= \lim_{k \to \infty} \left( \frac{2}{k+1} - \frac{2}{k(k+1)} \right) \\
&= \frac{2}{\infty + 1} - \frac{2}{\infty(\infty + 1)} \\
&= 0 - 0 \\
&= 0.
\end{aligned}$$

Similarly, using (22), (16), and (17), we have

$$\begin{aligned}
\lim_{k \to \infty} (w_1^p - w_e) &= \lim_{k \to \infty} (w_e - w_k^p) \\
&= \lim_{k \to \infty} \frac{1}{k} \frac{k-1}{k+1} \\
&= \lim_{k \to \infty} \left( \frac{1}{k+1} - \frac{1}{k(k+1)} \right) \\
&= \frac{1}{\infty + 1} - \frac{1}{\infty(\infty + 1)} \\
&= 0 - 0 \\
&= 0.
\end{aligned}$$

∎

As a result, the difference between the weights of the first and the last authors in the positional weight assignment scheme decrease as a function of the number of authors of a paper (Figure 2). The same applies to the difference between the weight assigned to each author in equal weight assignment scheme and that of the weights of either of the first author or of the last author in positional weight scheme (Figure 3).

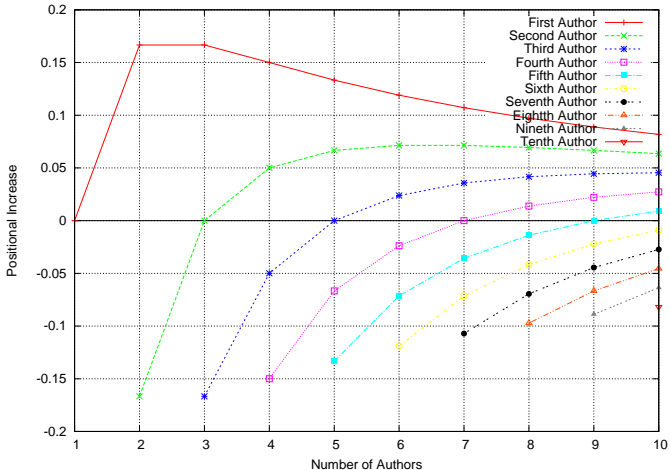

Fig. 4. Increase in the positional weights of authors as a function of the number of authors.

The expression in Lemma 3 says that the amount of increase in the positional weights of some authors of the paper is equal to the amount of decrease in the positional weights of the remaining authors. This is exemplified in Figure 4, which shows an increase in the positional weights (i.e. $w_j^p - w_e$) of authors as a function of the number of authors of the paper. Note that in Figure 4, the point lying along the vertical grid lines show an increase if the point lies above the *zero x-axis* horizontal line, and it shows a decrease if the point lies below the *zero x-axis* horizontal line. We observe that when there is only one author of a paper, the positional and equally assigned weight are the same. This results in 0 increase in the positional weight as compared to the equally assigned weight of the first author in a single authored paper. When the number of authors is 2, the increase in the positional weight of the first author is 0.1666, the same is the decrease in the positional weight of the second author. In general, if there are $k$ authors in a paper, the amount of increase in the positional weight of 1st author will be equal to the amount of decrease in the $k$th author of the paper. Consider, for example, the number of authors of the paper to be equal to 5. The amount of increase in the positional weight of 1st author is equal to the amount of decrease in the 5th author, and the amount of increase in the positional weight of 2nd author is equal to the amount of decrease in the 4th author; the amount of increase/decrease in the 3rd author is 0. The same is true for the remaining numbers of authors.

## IV. RESULTS AND DISCUSSION

We first compare the proposed *positional indices* with the *h-index* through an example and then present results for a set of authors.

### A. Comparison of Weighted Indices with H-Index

In this section, we compare the proposed weighted indices with an existing index known as $h$-index [1]. As mentioned earlier, $h$-index is simple to compute, however, its demerits are as follows.

- *Multiplicity of Authors*: It does not take into account multiplicity of authors of papers that belongs to the set of papers for $h$-index. Also, it does not take into account the contribution of authors to their individual papers. In other words, there can be two authors with the same $h$-index, however, most of the papers of one author may have multiple authors and the other author can be rather independent i.e. in most of his/her papers, he/she is either the sole author or the number of authors is much less than the former one. Depending upon the number of authors in each paper that belong to $h$-core, their individual contributions in each paper may not be the same.
- *Author Position*: It does not consider the order of the authors in the list of authors of each paper belonging to $h$-core. In other words, among the two authors with the same $h$-index, it may happen that one is the first author in most of his/her papers, and the other is the second or later author in most of his/her papers. Although, the contributions of the two authors to their individual papers are different, $h$-index does not reflect this difference.
- *Citation Spikes*: It does not take into account the spikes in citations of the papers belonging to the $h$-core. Suppose, the two authors have the same $h$-index, however, out of the papers forming $h$-core, some papers of one author have a large number of citations, and other papers have the number of citations just greater than or equal to $h$. On the other hand, the number of citations of the papers of the other author are just greater than or equal to $h$. Although, there is a significant difference between the quality of research produced by the two authors, however, $h$-index does not reflect this difference.

To understand it better consider the following example.

*Example 1:* The citations, number of authors, and author position of authors $A$ and $B$ are given in Table II[3]. Let us compute their $h$-indices and the proposed weighted indices.

*Solution*: We compute the $h$-indices and one of the proposed indices as follows.

- For computing the $h$-indices, note that 10 papers of both authors have at least 10 citations and the remaining $(12 - 10) = 2$ papers have less than 10 citations, therefore, $h$-indices of both the authors are 10.
- Let us now compute their weighted indices (specifically, we compute one of the weighted indices called the *positionally weighted citation h-cut* as defined in (9) of both the authors. The *positionally weighted citation h-cut*

---

[3]We consider this example with relatively a smaller $h$-index due to space limitations and just to illustrate the ideas. One can create an example with relatively a larger $h$-index.

TABLE II
*Example 2*: PAPER NUMBER, NUMBER OF CITATIONS, NUMBER OF AUTHORS, AND THEIR POSITION IN THE LIST OF AUTHORS (FICTITIOUS EXAMPLE).

| Paper Number | Number of Citations | | Number of Authors | | Author Position | |
|---|---|---|---|---|---|---|
| | A | B | A | B | A | B |
| 1 | 1048 | 34 | 2 | 2 | 1 | 1 |
| 2 | 997 | 32 | 1 | 2 | 1 | 2 |
| 3 | 886 | 27 | 1 | 2 | 1 | 2 |
| 4 | 797 | 23 | 2 | 1 | 2 | 1 |
| 5 | 665 | 21 | 1 | 2 | 1 | 2 |
| 6 | 623 | 18 | 1 | 2 | 1 | 2 |
| 7 | 546 | 17 | 2 | 1 | 2 | 1 |
| 8 | 15 | 15 | 3 | 2 | 3 | 2 |
| 9 | 12 | 13 | 2 | 1 | 2 | 1 |
| 10 | 10 | 10 | 1 | 1 | 1 | 1 |
| 11 | 8 | 7 | 2 | 1 | 2 | 1 |
| 12 | 7 | 6 | 1 | 1 | 1 | 1 |

of author $A$ is given by

$$\xi_p = 1048 \times \frac{1}{3} + 997 \times 1 + 886 \times 1 + 797 \times \frac{1}{3}$$
$$+ 665 \times 1 + 623 \times 1 + 546 \times \frac{1}{3} + 15 \times \frac{1}{3}$$
$$+ 12 \times \frac{1}{3} + 10 \times 1$$
$$= 3982.$$

The *positionally weighted citation h-cut* of author $B$ is given by

$$\xi_p = 34 \times \frac{2}{3} + 33 \times \frac{1}{3} + 27 \times \frac{1}{3} + 23 \times 1$$
$$+ 21 \times \frac{1}{3} + 18 \times \frac{1}{3} + 17 \times 1 + 15 \times \frac{1}{3}$$
$$+ 13 \times 1 + 10 \times 1$$
$$= 123.67.$$

Note that there is a large difference between the quality of research produced by author $A$ and author $B$. In general, the proposed weighted indices, specifically with positional weights, take into account not only the number of authors of each paper belonging to the $h$-core, they also take into account the position of the given author in the list of authors of each paper. As a result, the proposed weighted indices are able to differentiate among the two researchers based on the number of authors and their positional contributions to each of their papers belonging to their individual $h$-cores. Further, as opposed to $h$-index, the weighted indices take into account the spikes in the number of citations. In other words, they are able to differentiate among the two authors based on the weighted citations of each paper belonging to their respective $h$-cores.

*B. Performance of Indices*

We computed proposed indices for a set of authors with the following values of $h$-index: $\{84, 73, 67, 62, 58\}$. We used a freely available indexing database, called *Microsoft Academic Search* [18], for citations corresponding to authors with the given $h$-indices, specifically, of *networks and communications* group. For computing the positional indices, we followed the following procedure. Let $c_i$, $i = 1$ to $n$, be the citations of $i$th paper of the given author with $k$ number of authors and let $j$, $1 \leq p \leq k$, be the position of the author in the list of authors of the paper. We computed weighted citations $ci' = c_i w_i$ for each paper of the author. The assignment of weights, $w_i$, is discussed in the previous section. We then computed the indices for both position based weights as well as equal weights. We computed the positional $h$-index, $h_p$ from the weighted citations $c_i'$. There are two weighted $h$-indices: $h_p$ and $h_e$ depending upon which weighted citations we consider in computing them i.e. positionally weighted or equally weighted.

TABLE III
A COMPARISON AMONG $h$-INDEX, WEIGHTED $h$-INDICES, AND WEIGHTED CITATION $h$-CUTS FOR A GIVEN SET OF AUTHORS.

| Author | $h$-index | $h_p$ | $h_e$ | $\xi_w$ | $\xi_e$ |
|---|---|---|---|---|---|
| $A$ | 84 | 38 | 43 | 4568.91 | 6643.70 |
| $B$ | 73 | 34 | 40 | 4109.18 | 5183.72 |
| $C$ | 67 | 31 | 36 | 3469.41 | 4242.26 |
| $D$ | 62 | 28 | 33 | 2634.85 | 3638.79 |
| $E$ | 58 | 32 | 33 | 3724.68 | 4802.81 |

Table III shows the values of $h$-index, positionally weighted $h$-index, $h_p$, equally weighted $h$-index, $h_e$, positionally weighted citation cut, $\xi_p$, and equally weighted citation cut, $\xi_e$. Authors are listed in decreasing order of their original $h$-index. This forms one ranking of authors considered herein, namely $\{A, B, C, D, E\}$, in that order, which is based on the original $h$-index. However, a closer look reveals that even though the original $h$-index of author $E$ (with rank 5) is the least, however, his/her positionally weighted $h$-index $h_p = 33$ is better than the authors $D$ whose positionally weighted $h$-index is $h_p = 28$. Based on positionally weighted $h$-index, the rank of author $E$ is ahead of author $D$ i.e. the new ranking is $\{A, B, C, E, D\}$, in that order. This fact is visible from Figure 5, which shows $h$-index, positionally weighted $h$-index, and equally weighted $h$-index for authors numbered 1 (or $A$) through 5 (or $E$). On the basis of equally weighted $h$-index, $h_e$, authors $D$ and $E$. However, equally weighted $h$-index, $h_p$, of author $D$ and $E$ is the same and is equal to 33. Therefore, on the basis of equally weighted index, $h_e$, the ranking of authors is $\{A, B, E, \{C, D\}\}$.

On the other hand, on the basis of weighted citation $h$-cuts, $\xi_w$ and $\xi_e$, author $E$ is better than authors $D$ and $C$, which have ranks 4 and 3, respectively. This fact is clearly visible from Figure 6, which shows a comparison of weighted citation $h$-cuts of authors 1 (or $A$) through 5 (or $E$). As a result, based on the weighted citation $h$-cuts, which consider the number of authors and the position of authors, the ranking of the given set of authors needs to be modified. The new ranking, then, is $\{A, B, E, C, D\}$, in that order.

Figure 7 shows the number of citations received by the top $h$ papers of the given set of authors (i.e. from author numbered $A$ to author numbered $E$) in descending order of the number of citations received by individual papers. Note that the number of citations of each author decrease with the paper number because the papers are arranged in descending order of their citation counts. Also, we observe that the number of citations of some authors, such as author $A$ and author $E$, are very large for the first few papers as compared to the papers that appear later in the list of cited papers. Specifically, the number of citations of author $E$ for the first few papers are larger than that of author $A$, however, the later papers of author $E$ are cited less as compared to author $A$. As mentioned earlier, $h$-index of author $A$ is larger than that of author $E$ (see Table III), therefore, in the ranking which is based on $h$-index, author $A$ precedes author $B$. However, the citations of first few papers of author $E$ are larger than those of author $C$ and author $D$, and for later papers citations of author $E$ are comparable to those of authors $C$ and $D$, therefore, the weighted citations of author $E$ are larger than those of author $C$ and author $D$. The reason is that the first few papers of author $E$ enhance the number of papers with larger number of weighted citations as compared to author $C$ and author $D$. Consequently, the weighted $h$-index of author $C$ is larger than those of author $C$ and author $D$ (see Table III).

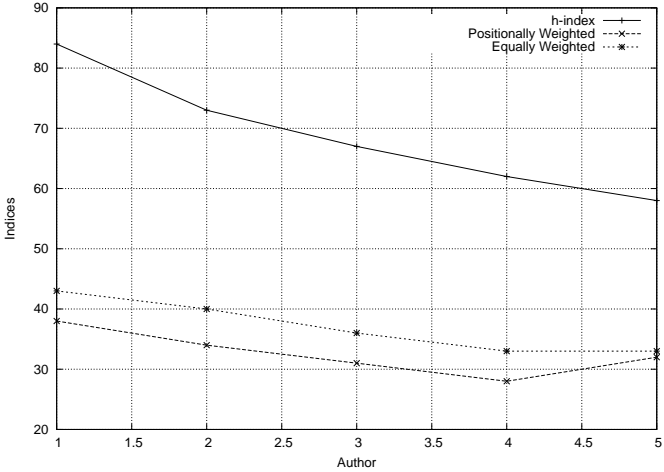

Fig. 5. Comparison among $h$-index, positionally weighted $h$-index, $h_p$, and equally weighted index, $h_e$ for a given set of authors.

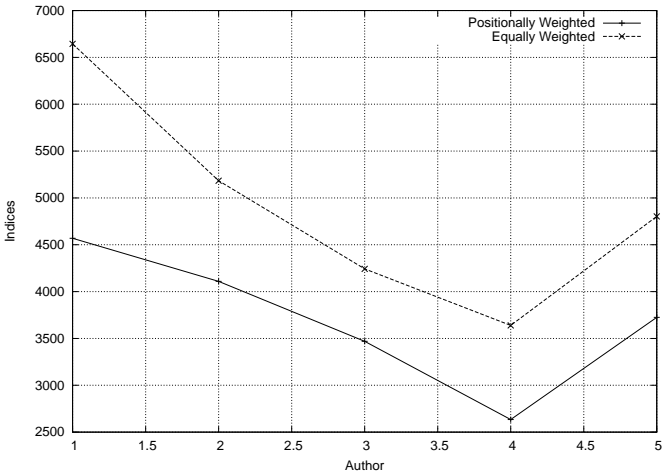

Fig. 6. Comparison between weighted citation $h$-cuts: positionally weighted $\xi_p$, and equally weighted $\xi_e$ for a given set of authors.

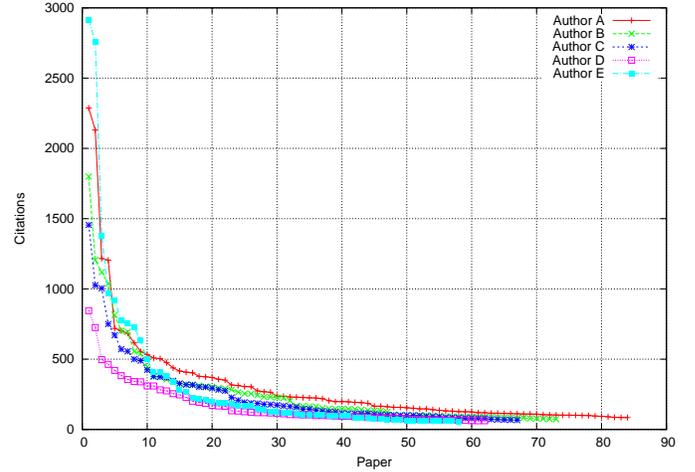

Fig. 7. Citations of the top $h$ papers of a given set of authors.

Figure 8 shows positionally weighted citations of the top $h$ papers of a given set of authors. Note that in case of positionally weighted citations, the citations of a paper are divided among the authors of the paper according to their positional weights, and the positional weight assigned to a given author depends upon the number of authors and the position of the given author in the list of authors of the paper. Comparing this with Figure 7, we observe that the descending order in the number of citations of individual papers as in Figure 7 is not maintained in Figure 8. The reason is that the number of authors and the position of the given author varies from paper to paper. As the weights assigned depend on the number of authors and the position of author, therefore, the weights vary from paper to paper. As a result, the positionally weighted citations vary from paper to paper. Therefore, the original order among the citations of the papers may not be preserved.

Figure 9 shows equally weighted citations of the top $h$ papers of a given set of authors. Again, we wish to emphasize that the descending order in the number of citations of individual papers may not be maintained in case of equally weighted citations. Although, in equally weighted citations, the citations of a paper are divided equally among the authors of the paper. However, the number of authors of each paper that are arranged in descending order of the number of citations are different. Therefore, the number of citations after dividing them with the corresponding number of authors may not be in the descending order. This accounts for the observed behaviour.

## V. COMPARISON WITH THE RELATED WORKS

A modification of $h$-index, called $h_m$ index, that takes into account multiple co-authors, is proposed in [19], [20]. Another variant of $h$-index that takes into account multiple authors, called an *adapted pure h-index*, is proposed in [21]. Let us denote the adaptive pure $h$-index by the symbol $h_a$ where the subscript $a$ is for the qualifier "adaptive". In adaptive pure $h$-index, the number of citations of a paper are divided by the square-root of the number of authors of the paper.

A mathematical theory of the $h$-index and $g$-index in case of fractional counting of authorship is described in [22]. Therein, citation counts of $i$th paper of an author are divided by the number of authors of paper $i$, and then variants of $h$-index and $g$-index, called fractional $h$-index, $h_f$, and fractional $g$-index, $g_f$, respectively, are computed. Another variant of the fractional $h$-index called $h_F$ index (and correspondingly the variant of $g$index called $g_F$ index, however, in this paper we are interested only in $h$-index and not the $g$-index) is obtained such that the summation of the inverse of the number of authors is greater than or equal to the the summation of the top citation counts (provided that the citation counts are arranged in the descending order). The idea behind computing the fractional $h$-index, specifically computing $h_F$, is novel and applies well when one considers the publications of an author. However, it logically seems to isolate or decouple the number of citations from the number of authors of the individual papers while comparing a number of authors. For example, consider that there are two authors $A$ and $B$, $j$th papers of $A$ and $B$ have received $a$ and $b$ number of citations, respectively, and have the same number of authors, say $m$. Irrespective of the actual values of $a$ and $b$, there is an increment of $\frac{1}{m}$ in the fractional $h$-index, $h_F$. This is justified when the values of $a$ and $b$ are comparable, however, if there is a large difference between the values of $a$ and $b$, then not taking into account their actual values in the computation of the index does not seem to be justified. For example, if $a = 10$, $b = 100$, and $m = 3$ will incur an increment of $0.333$ in $h_F$ for both the authors, irrespective of how large the difference is between the number of citations of both the authors for $j$th paper. Comparing our weighted indices to the fractional index, $h_F$, proposed in [22] reveals that the weighted indices take into account the weighted citations of individual papers of each author.

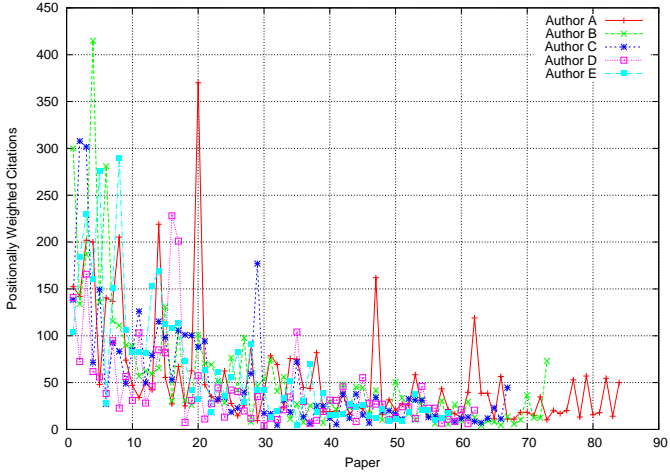

Fig. 8. Positionally weighted citations of the top $h$ papers of a given set of authors.

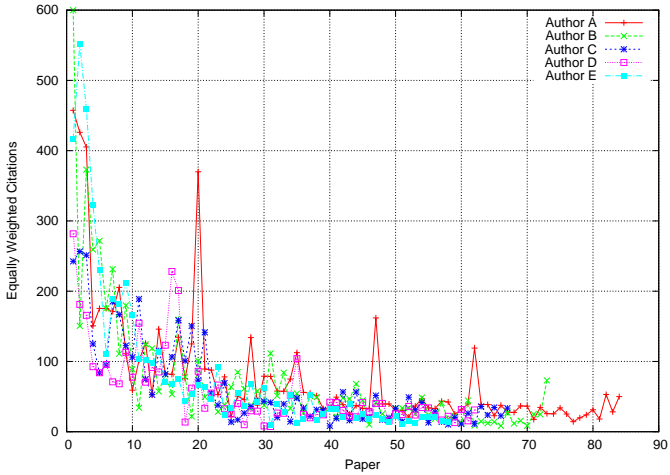

Fig. 9. Equally weighted citations of the top $h$ papers of a given set of authors.

To account for the multiple authors of papers, a harmonic allocation of authorship credits is described in [23] in the context of what kind of biases one may have for authors of the paper. The harmonic allocation of authorship credits was suggested in [24] and recently in [25]. It is discussed in [26] that harmonic weights enable authors to share authorship credits equitably rather than equally. Therein, it has been discussed that harmonic authorship credit scores can be applied to authors in different fields such as psychology, medicine, chemistry. Further, four types of authorship credits—harmonic, arithmetic, geometric, fractional— are discussed in [26].

Our concept of positional weights is similar to the harmonic allocation of authorship credits proposed in [23]. However, in this paper, we have carried out a rigorous mathematical analysis of the positional weights and we have derived the expressions related to the positional weights. Further, we have discussed that the $h$-index with positional weights, $h_p$, would be more relevant as compared to the original $h$-index, because $h_p$ takes into account multiple authors and their relative contributions in the papers belonging to the $h$-core. We have pointed out that when one arranges the number of citations of papers in the descending order, the $h$-core when the citations are not weighted may not be the same as the $h$-core when the citations are weighted. Note that the weighted $h$ indices, be it positionally weighted or equally weighted, do take into account multiple authors in the papers, however, they do not reflect the spikes in the number of weighted citations. To take into account the spikes in the number of weighted citations of an author, we defined another class of indices called *weighted citation $h$-cuts*. To explain how to compute *weighted $h$-indices* and *weighted citation $h$-cuts*, we considered a fictitious example and an example from the real world—a group of authors belonging to the *network and communication* from a freely available indexing database *Microsoft Academic Search*. Further, we would like to mention that our concepts of *weighted $h$-indices* and *weighted citation $h$-cuts* are not limited to only positional weights, rather, they are open to *any* kinds of weights that seem to be appropriately share authorship credits. This is because we have taken into account the *policy* according to which the weights are assigned in the definitions of indices. A change of the policy for assigning the weights shall incur a corresponding change in the weights for the authorship, and the indices shall change accordingly.

On the other hand, an index to quantify the scientific research output of an individual author that takes into account the effect of multiple co-authorship proposed in [28]. The index is called $\hbar$-index (pronounced as *hbar*-index), and is defined as the number of papers of an individual author having citation counts larger than or equal to the $\hbar$ of all co-authors of each paper. Table IV shows a comparison and the relative features of the $h$-index and its variants together with the indices proposed in this paper. Note that in Table IV, we consider only those indices which take into account multiple authorship except the original $h$-index.

TABLE IV
A COMPARISON PROPOSED WEIGHTED INDICES WITH THE INDICES THAT TAKE INTO ACCOUNT MULTIPLE AUTHORSHIP.

| *Index* | *Basis* | *Features* | *Comments* |
|---|---|---|---|
| $h$-index [1] | $\min\{c_i\} \geq h$ | Simple to compute | Does not consider individual contributions |
| Adaptive $h$-index [21], $h_a$ | $c_i = \frac{c_i}{\sqrt{a_i}}$ $\min\{c_i \geq h\}$ | Square root normalization of authors | |
| Modified $h$-index [19], $h_m$ | $r(i) = \sum_{i=1}^{r} \frac{1}{a_i}$ $c(r(h_m)) \geq h_m$ $\geq c(r(h_m)+1)$ | Favours authors with a flat frequency of citations | Excess citations not counted |
| Fractional $h$-index [22], $h_f$ | $\sum_i \frac{1}{a_i} \geq \sum_i c_i$ | Fractional counting | Excess citations not counted |
| $\hbar$-index [28] | $p_i : c_{p_{i,j}} \geq \hbar, \forall a_j$ of $p_i$ | Favours senior coauthors | Excess citations not counted |
| Weighted $h$-indices, $h_w$, ($h_p$ and $h_e$) | $\min\{w_i c_i\} \geq h$ | Open to any weight assignment | Excess citations not counted |
| Weighted Citation Aggregate, $\psi_w$, ($\psi_p$ and $\psi_e$) | $\sum_i w_i c_i, \quad w_i \in \mathcal{A}$ | - do - | Excess citations counted at both ends |
| Weighted Citation $h$-cut, $\xi_w$, ($\xi_p$ and $\xi_e$) | $\sum_{i=1}^{\hbar} c_i w_i$ | - do - | Excess citation counted at upper end |

However, $h$-index and its variants such as $h_m$ [19], $h_a$ [21], $h_f$ [22], $\hbar$ [28], including the weighted $h$-index proposed in this paper do not take into account the portions of citations that are greater than or in excess of the corresponding $h$-index or its variants. It is to note that there is something significant to be revealed, about the quality of research produced by an author, by the portion of citations that is in excess to those needed to compute the $h$-index or its variants and should not simply be ignored. Therefore, there is a need to have a supplement to the $h$-index or its variant even for multi-authored papers. One of the supplement can be the $g$-index proposed in [16]. The $g$-index considers the missing upper and lower portions of citations of authors. However, while having an $h$-index, one would not like to have the missing lower citations of an individual because that is not going to enhance the quality of publications. In fact, it is the upper portion or the excess in number of citations belonging to the corresponding $h$-core of the index that may affect (or enhance) the quality of publications of an individual. We would like to mention that the *weighted citation $h$-cuts* may serve as a supplement to the $h$-index or its variants as it does take into account the number of citations that are in excess of that required to compute the index.

## VI. CONCLUSION

The design of an index to quantify the quality of research is a challenging task. In this paper, we proposed a set of indices for determining the quality of research, namely, *weighted citation aggregate*, *weighted h-index*, and *weighted citation h-cuts*. Our indices try to take into account the contributions made by an author in a multi-authored paper. As there is no mechanism to exactly determine the contribution of an individual author in a multi-authored paper, therefore, we assumed that the position of an author in the list of authors of the paper may provide an indication about their individual contributions, unless specified explicitly. In other words, the contribution of an author is assumed to be in accordance with the order in which the names of authors appear in the list of authors of the paper. Authors whose names appear earlier in the list are assumed to have a larger contribution as compared to the authors whose names appear later in the list. We have used two weight assignment policies: *positionally weighted* and *equally weighted*. We have analyzed these weight assignment policies and compared their effects on the weighted indices. We would like to mention that the definitions of our weighted indices are open to any weight assignment policy or scheme that seems to address multiple authorship in an appropriate manner. We compared our indices with the $h$-index for a selected set of authors. As opposed to $h$-index, a class of index that we called the *weighted h-indices* take into account the weighted contributions of authors in multi-authored papers. Further, the other class of weighted indices that we called *weighted citation h-cuts*, take into account the number of citations that are in excess to determining an index, and may serve as a supplement to $h$-index or its variants. In future, one may study the effect of citation contexts on the quality of research produced by an author.